# Super-enhancement focusing of Teflon sphere

Liyang Yue[1*], Zengbo Wang[1*], Bing Yan[1], James N. Monks[1], Yasir Joya[1,2], Rakesh Dhama[1], Oleg V. Minin[3], and Igor V. Minin[3*]

[1]School of Computer Science and Electronic Engineering, Bangor University, Dean Street, Bangor, Gwynedd, LL57 1UT, UK

[2]Faculty of Materials and Chemical Engineering, Ghulam Ishaq Khan Institute of Engineering Sciences and Technology, Topi, Khyber Pakhtunkhwa, 23640, Pakistan

[3]National Research Tomsk Polytechnic University, Lenin Ave., 30, Tomsk, 634050, Russia

Email:   Dr Liyang Yue, l.yue@bangor.ac.uk
         Dr Zengbo Wang, z.wang@bangor.ac.uk
         Prof Igor Minin, prof.minin@gmail.com

**Abstract**

A Teflon (Polytetrafluoroethylene, PTFE) sphere can be used as a focusing lens in the applications of imaging and sensing due to its low-loss property in the terahertz (THz) band. In this paper, we analytically calculated field intensities and focusing parameters for Teflon spheres at different low-loss levels and then discovered a super-enhancement focusing effect in the spheres with particular size parameters, which can stimulate about 4000 times stronger field intensity than that for incident radiation as well as the great potential of overcoming diffraction limit despite high sensitivity to the magnitude of Teflon loss. A subsequent analysis of scattering amplitudes proved that the strong scattering of a single order mode in the internal electric or magnetic field is the main factor causing this phenomenon.

## 1. Introduction

Lord Rayleigh stepped forward to scattering study of small particles and found that a small spherical dielectric particle with a radius - $a$, which is much smaller than the wavelength of incident light - $\lambda$ ($a/\lambda \ll 1$), can symmetrically scatter light in the forward and backward directions like a point electric dipole [1]. Lorenz-Mie theory fully described the optical absorption and scattering of light through a homogeneous sphere with a size that was similar to the wavelength of incident light ($a/\lambda \cong 1$) thereafter and involved spherical multipole partial waves to solve the Maxwell equations [2]. Besides, an asymmetrical jet-like near-field focus situated near the shadow surface of a dielectric sphere sized in a certain range of $a/\lambda$ (mesoscale dimension, $1 < a/\lambda < 30$) [3-5] has been well studied with a popular name 'photonic jet' since 2000 [6-9]. Its typical features include strong enhancement of field intensity, little divergence for the propagation of several wavelengths, transverse dimension smaller than diffraction limit, etc. [9]. Many approaches, e.g. metalens, assembly of nanofibres, and addition of a pupil-mask [10-13], were carried out to further improve these features for the photonic jet, but most of them were based on the geometrical or morphological changes of the sphere and difficult to realise using a relatively easy experimental setup.

Meanwhile, lossless ideal materials were used in the majority of theoretical research of dielectric particle focusing, which might be acceptable and less influential to the materials working in the optical band, however, there is a different criterion on that for the materials used in the applications of THz and millimetre wave radiation. In fact, a material possessing a loss tangent – $\tan\delta$ at the order of magnitude of 1e-4 can be called a low-loss material in the THz band [14]. The Mie resonance effect and focusing characteristics of a sphere made from a two-component $TiO_2$-PE (Polyethylene) composition with a $TiO_2$ volume fraction of 0.75 and the effective refractive index - $n$ = 2.5 and extinction coefficient - $k$ = 1.985e-2 were numerically studied in the wavelength range of 150 - 600 μm (THz band) by Storozhenko et al. [15], but this material is difficult to synthesise and expensive. Teflon is such a kind of material with an extremely small extinction coefficient – $k$ in the wavelength range of 30 μm to 1 mm or the frequency range of 300 GHz to 10 THz. The lenses and probes made of it have not only the strength of low loss but also the advantages on cost and focusing capability compared to those from composite dielectric materials and noble metals.

In this paper, we extend our research scope to investigate focusing of the Teflon spheres with multiple size parameters determined from Lorenz-Mie theory while taking into account loss impact. Minin et al. studied similar focus in THz band by a non-resonant dielectric cuboid particle and found it can be formed in both transmission and reflection modes [16, 17]. The characteristics of resonance modes were investigated in a low-loss high-index dielectric sphere made of alumina ceramic ($e_r$ = 10.7 and $\tan\delta$ = 0.9e-3) to highlight high-gain and high radiation efficiency achieved in an antenna design based on it by Dey and Hesselbarth [18]. Yue et al. and Wang et al. studied focusing of specifically sized dielectric spheres in their publications and they thought the unusual Poynting vector circulation and Fano resonance could result in the extraordinary near-field focus with a large field intensity by a high-index sphere [19, 20]. However, these researchers did not consider the difference of focusing parameters for the individual spheres with the neighbouring size parameters and a relatively low refractive index, also the material intrinsic loss in their approaches. For filling this gap we created an analytical algorithm to calculate the key focusing parameters of Teflon spheres, including peak electric field $|E^2|$ intensity, transverse dimension in full width at half maximum (FWHM), and focal position, through scanning of size parameters of Lorenz-Mie theory with a variety of experimentally measured loss magnitudes of Teflon [21-23]. A super-enhancement focusing effect was discovered in the spheres with the particular size parameters in this process. The field intensity distributions of these spheres and the corresponding scattering amplitudes of each order mode were demonstrated in this paper to investigate its formation mechanism.

## 2. Results
### 2.1 Model
The algorithm used in this study was coded by both FORTRAN and MATLAB to calculate the focusing parameters of Teflon spheres with successive size parameters in a range. The size parameter of the Lorenz-Mie theory – $q$ can be expressed as [24],

$$q = \frac{2\pi a}{\lambda} \qquad (1)$$

where $a$ and $\lambda$ are Teflon sphere radius and wavelength of the incident light, respectively. The configurated scanning range of size parameter - $q$ was from $\pi$ to $20\pi$ with a step size of 0.1. A plane wave radiation polarised in $x$-axis direction was designed to illuminate the Teflon sphere along the $z$-axis from negative to positive. Values of $n$ and $k$ in the algorithm are treated as constants to simplify the relative permittivity – $\varepsilon$ of Teflon in THz band ($\varepsilon = n + k\mathrm{i}$ where $n$ and $k$ denote refractive index and extinction coefficient respectively) and the background medium is the air with $n = 1$. Here refractive index of Teflon was defined as 1.43 [25] and its extinction coefficient referring to several experimentally measured values which were 1.4e-4 [21], 5.95e-4 [22] and 1.7e-3 [23]. An ideal case of lossless Teflon with $k = 0$ was also calculated by the algorithm as a comparison. The algorithm spatially scanned the central plane of the sphere within the range of -1.5$a$ to 1.5$a$ in order to collect the data of peak field enhancement, focus position, and transverse dimension in an individual sphere model. The number of data collection points was set to 500 per $a$ in a sphere and its vicinity, which meant that a total of 1500 points were sampled with a specific size parameter. Besides, the transverse dimension of a focus was measured in FWHM in both transverse electric and magnetic modes (TE and TM modes) and denoted by $FWHM_{TE}$ and $FWHM_{TM}$ in comparison with the normalised Rayleigh criterion ($RC$) and diffraction limit ($DL$) in Teflon ($n = 1.43$) to the sphere radius – $RC/a$ and $DL/a$, which are written in the function of $q$ based on Eq. (1) as [26],

$$\frac{RC}{a} = \frac{0.61\lambda}{a} = 0.61\frac{2\pi a}{qa} = \frac{1.22\pi}{q} \qquad (2)$$

$$\frac{DL}{a} = \frac{\lambda}{2na} = \frac{2\pi a}{2naq} = \frac{\pi}{nq} = \frac{\pi}{1.43q} \qquad (3)$$

**2.2 Super-enhancement**

The statistics of all focusing parameters are illustrated as the functions of $q$ with the curves of loss levels of $k = 0$, 1.40e-4, 5.95e-4, and 1.7e-3 in Figure 1. Figure 1 (a) summarises the $|E^2|$ peak field enhancements for all spheres and they manifest an increasing tendency with the regular oscillations for all 4 curves. Several super-enhancement giant peaks appear in the curves of $k = 0$ and 1.40e-4 at the particular $q$ values, especially for the ideal case of lossless Teflon with $k = 0$. It has the most super-enhancement $q$ positions and the highest peak with a maximum $|E^2|$ field enhancement of 3950 presents at the position of $q = 28.64159$. This magnitude is much larger than that reported in the previous literature investigating Teflon lens focusing. Another similar super-enhancement peak is at the position of $q = 46.54159$ and reaches $|E^2|$ field intensity of 2952. The increase of Teflon loss level effectively weakens this super-enhancement, which leads to two less fluctuant curves of $k = 5.95$e-4 and 1.70e-3 in Figure 1 (a). Figure 1 (b) exhibits all focal positions normalised to the particle radius – $a$ (distance between the focus and the sphere centre divided by the sphere radius) with a dashed line at the position of 1.0 referring to the lower boundary of the sphere. The majority of the markers are above the dashed line of position 1.0, which means these focuses are located below the lower boundary and outside the sphere. By contrast, it is noted that a few markers of the cases of $k = 0$ and $k = 1.40e-4$ are below the dashed line of position 1.0 and in the shaded area. The size parameters of these markers also correspond to the $q$ positions of the super-enhancement giant peaks shown in Figure 1 (a), which proves that all super-enhancement focuses are inside the sphere and mainly distributed in the area between position 0.8 and position 1.0 across the $z$-axis, as shown in Figure 1 (b).

Figure 1 (c) and (d) demonstrated the normalised transverse dimensions of the focuses to sphere radius in FWHM as a function of $q$ for TE mode (FWHM$_{TE}$/$a$) and TM mode (FWHM$_{TM}$/$a$), respectively. Here we compare these dimensions to the normalised Rayleigh criterion – $RC/a$ and diffraction limit – $DL/a$ as well (based on Eq. (2) and (3)), then display them as two dashed curves in Figure 1 (c) and (d). It is demonstrated that FWHM$_{TE}$/$a$ and FWHM$_{TM}$/$a$ have a general relationship with the Rayleigh criterion as FWHM$_{TE}$/$a$ < $RC/a$ < FWHM$_{TM}$/$a$, which results in an elliptical focus profile with a major axis in $x$-axis direction and is in accordance with Heisenberg's uncertainty principle applied in the scattering of photons [27]. However, this tendency cannot apply to the spheres stimulating super-enhancement effect. Several of their FWHM$_{TE}$/$a$ can overcome the diffraction limit, as shown in Figure 1 (c), and offer the great potential of approaching or overcoming the diffraction limit for these spheres.

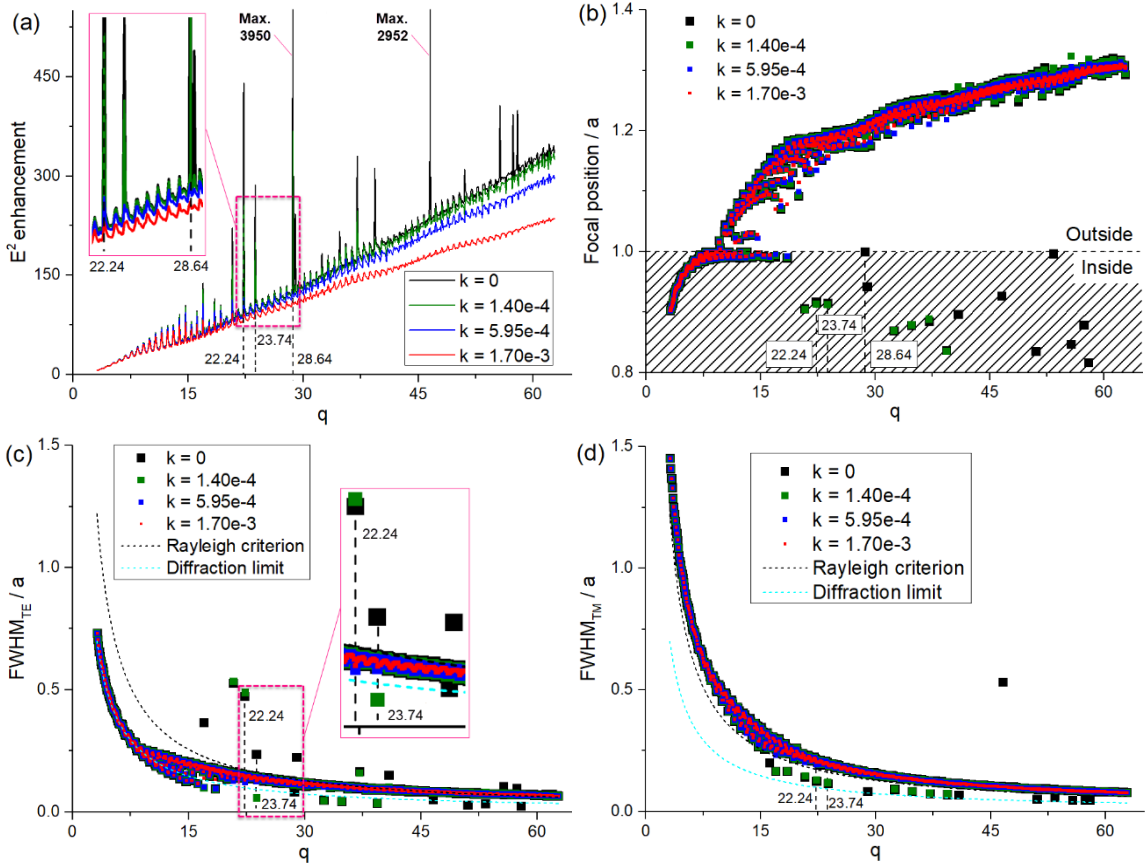

Figure 1. (a) $|E^2|$ peak field enhancements, (b) normalised focal lengths, (c) and (d) FWHM$_{TE}$ and FWHM$_{TM}$, in the function of $q$ with k = 0, 1.4e-4, 5.95e-4, and 1.70e-3.

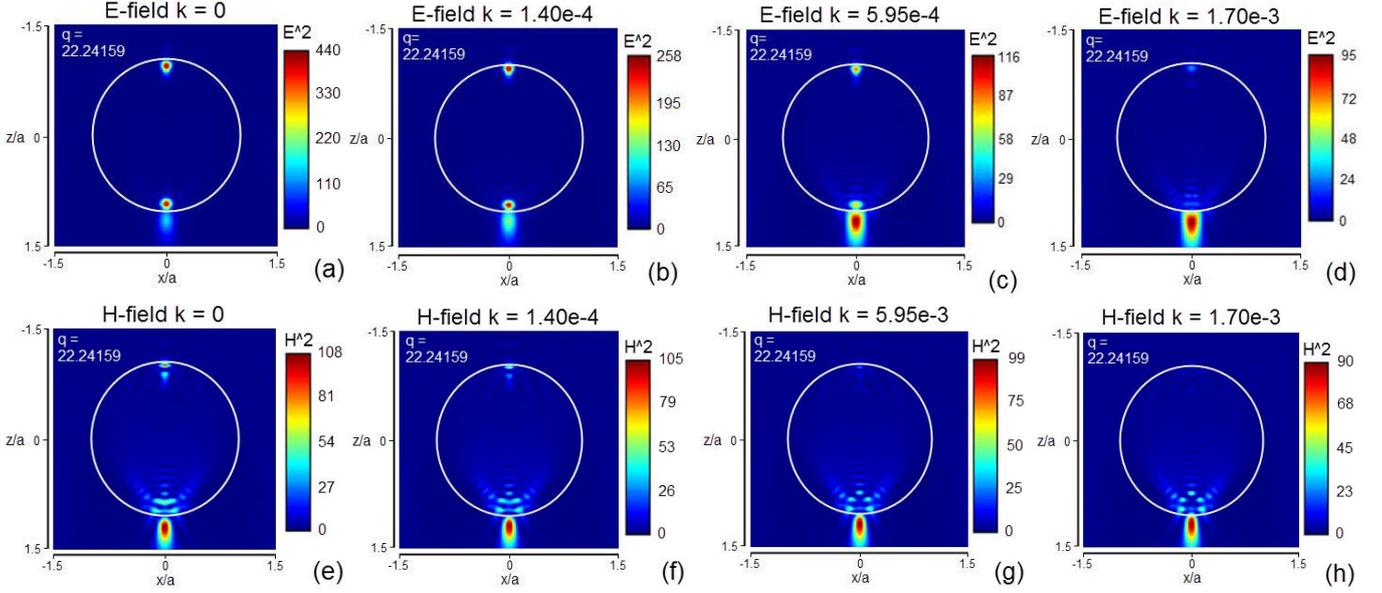

Figure 2. Distributions of $|E^2|$ (a-d) and $|H^2|$ (e-h) field intensities for the sphere with $q = 22.24159$ at the loss levels of $k = 0$, 1.40e-4, 5.95e-4, and 1.70e-3 in TM mode.

Figure 2 and 3 manifest the distributions of electric and magnetic field intensities ($|E^2|$ and $|H^2|$) on *xy* plane (TM mode) at 4 loss levels for two super-enhancement Teflon spheres with the size parameters of 22.24159 and 28.64159, respectively. Their positions in Figure 1 (a) are marked by two black dashed lines. It should be mentioned that the near-field distributions of the super-enhancement effect occurring in the spheres with $k = 0$ and 1.4e-4 are generally in the form of hotspots situated around the sphere poles in both electric and magnetic fields. The differences between these two super-enhancement examples are the shape and number of the hotspots around the pole and the layout in the magnetic field. For the sphere with $q = 22.24159$, the shape of hotspots is more circular, and the jet-shape focuses can be found in the magnetic field layouts besides the polar features. However, the sphere with $q = 28.64159$ has the longitudinally symmetric oval hotspots at the poles in both electric and magnetic fields pairing with the patterns of Whispering-gallery mode. Meanwhile, Figure 2 and 3 show that the increase of loss level can inhibit this super-enhancement effect, which reflects on the drop of the field intensities of the hotspots with the growth of *k* value, especially for the hotspot at the upper pole. Therefore, the super-enhancement of $|E^2|$ and $|H^2|$ field intensities only can be stimulated at the loss levels of $k = 0$ and 1.4e-4 with the existence of the upper polar hotspots in this study. At the loss levels of $k = 5.95$e-4 and 1.70e-3, the near-field distributions in two examples transition into the classic photonic jet layout, while decreasing peak enhancements. The near-field distributions of the normal spheres with the neighbouring *q* values of 22.14159 and 28.74159 are included in Figure S1 and S2 of the supplement of the paper. They present low field enhancements and typical photonic jets at all loss levels in both electric and magnetic fields.

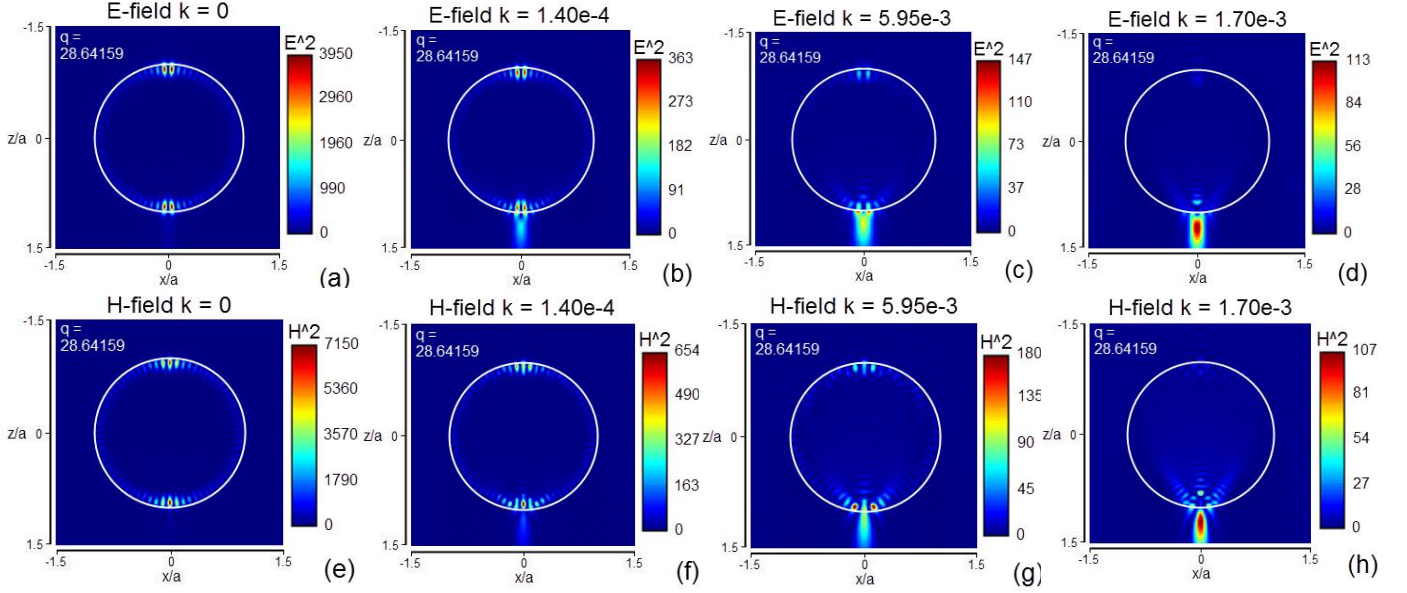

Figure 3. Distributions of $|E^2|$ (a-d) and $|H^2|$ (e-h) field intensities for the sphere with $q = 28.64159$ at the loss levels of $k = 0$, 1.40e-4, 5.95e-4, and 1.70e-3 in TM mode.

## 3. Discussion
### 3.1 Scattering amplitude

The regular oscillation of the peak field enhancement displayed in Figure 1 (a) is considered to be caused by the optical resonance in a spherical cavity [28]. In this case, the scattering of a plane wave radiation is a collective effect contributed by the independent order modes. The scattering amplitude of a single order mode is determined by the complex scattering wave coefficients – $B_l^e$ for electric field and $B_l^m$ for magnetic field [29]. These two coefficients are expressed as follows [29]:

$$B_l^e = i^{l+1} \frac{2l+1}{l(l+1)} \frac{\hat{n}\psi_l^{'}(q)\psi_l(\hat{n}q) - \psi_l(q)\psi_l^{'}(\hat{n}q)}{\hat{n}\zeta_l^{(l)'}(q)\psi_l(\hat{n}q) - \zeta_l^{(l)}(q)\psi_l^{'}(\hat{n}q)} \quad (4)$$

$$B_l^m = i^{l+1} \frac{2l+1}{l(l+1)} \frac{\hat{n}\psi_l(q)\psi_l^{'}(\hat{n}q) - \psi_l^{'}(q)\psi_l(\hat{n}q)}{\hat{n}\zeta_l^{(l)}(q)\psi_l^{'}(\hat{n}q) - \zeta_l^{(l)'}(q)\psi_l(\hat{n}q)} \quad (5)$$

where $l$ is the number of order mode, $\hat{n}$ is the complex refractive index of Teflon sphere relative to the surrounding medium - air, and $\psi_l(q)$ and $\zeta_l^{(l)}(q)$ are defined by [29],

$$\psi_l(q) = \sqrt{\frac{\pi q}{2}} J_{l+\frac{1}{2}}(q) \quad (6)$$

$$\zeta_l^{(l)}(q) = \psi_l(q) - i\chi_l(q) = \sqrt{\frac{\pi q}{2}} H_{l+\frac{1}{2}}^{(1)}(q) \quad (7)$$

$$\chi_l(q) = -\sqrt{\frac{\pi q}{2}} N_{l+\frac{1}{2}}(q) \quad (8)$$

where $J_{l+\frac{1}{2}}(q)$, $H_{l+\frac{1}{2}}^{(1)}(q)$, and $N_{l+\frac{1}{2}}(q)$ are the Bessel functions, the Hankel functions, and the Neumann functions, respectively [29]. The size parameter – $q$ and number of order mode - $l$ as the only two variables are highly involved with the complex functions as the denominators in the equations of $B_l^e$ and $B_l^m$ (Eq. (4) and (5)). Hence, these two variables with particular values can make the denominator in Eq. (4) – $B_l^e$ or Eq. (5) – $B_l^m$ very small, even approach 0, in the mathematical circumstance of a similar magnitude for the terms of $\hat{n}\zeta_l^{(l)'}(q)\psi_l(\hat{n}q)$ and $\zeta_l^{(l)}(q)\psi_l'(\hat{n}q)$, or the terms of $\hat{n}\zeta_l^{(l)}(q)\psi_l'(\hat{n}q)$ and $\zeta_l^{(l)'}(q)\psi_l(\hat{n}q)$, based on their subtractive relationship. This could explain the principle of the discovered super-enhancement effect and its strong dependence on the size parameter. Also, it is known that scattering amplitude of a single order mode decreases and approaches 0 with an increase of order mode number - $l$, and the number of that order mode with 0 contribution - $l_0$ can be expressed by size parameter - $q$. Its empirical formula is given by [30],

$$l_0 \approx q + 4.3q^{\frac{1}{3}} + 1 \tag{9}$$

The algorithm calculates the collective field intensity enhancement for every sphere using an iteration of all order modes until $l_0$ is reached. Consequently, field intensity enhancement of a sphere with a large size parameter (large $q$) should be larger than that for a sphere with a small size parameter (small $q$) due to more contributions from high order modes [31, 32], which leads to a general increasing tendency for all curves in Figure 1 (a). Surprisingly, the super-enhancement effect can break this tendency and let the sphere with a relatively small size parameter generate the largest field intensity enhancement within the scanning range, e.g. the sphere with $q$ = 28.64159 shown in Figure 1 (a) and Figure 3.

## 3.2 External and internal fields

According to the Lorenz-Mie theory, the collective scattering of radiation is spatially decided by both external and internal fields of a sphere (outside and inside of the sphere). This is represented by 4 factors of scattering amplitudes: $a_l$ for the external electric field, $b_l$ for the external magnetic field, $c_l$ for the internal magnetic field, and $d_l$ for the internal electric field [4]. Their formulas are expressed as follows:

$$a_l = \frac{F_l^{(a)}}{F_l^{(a)} + iG_l^{(a)}} \;,\; b_l = \frac{F_l^{(b)}}{F_l^{(b)} + iG_l^{(b)}} \tag{10}$$

$$c_l = \frac{in}{F_l^{(b)} + iG_l^{(b)}} \;,\; d_l = -\frac{in}{F_l^{(a)} + iG_l^{(a)}} \tag{11}$$

Meanwhile, $F_l^{(a,b)}$ and $G_l^{(a,b)}$ can be written in the Bessel and Neumann functions [4]:

$$F_l^{(a)} = n\psi_l'(q)\psi_l(nq) - \psi_l(q)\psi_l'(nq),\; G_l^{(a)} = n\chi_l'(q)\psi_l(nq) - \psi_l'(nq)\chi_l(q) \tag{12}$$

$$F_l^{(b)} = n\psi_l'(nq)\psi_l(q) - \psi_l(nq)\psi_l'(q),\; G_l^{(b)} = n\chi_l(q)\psi_l'(nq) - \psi_l(nq)\chi_l'(q) \tag{13}$$

where $\psi_l(q)$ and $\chi_l(q)$ are defined by Eq. (6) and (8).

Using Eq. (4) – (13), scattering contribution of each order mode can be quantified by these 4 factors and this information is useful to identify which field and order mode plays the most important role in the super-enhancement effect. Figure 4 demonstrates the scattering amplitudes of all 4 factors against order mode – $l$ for the super-enhancement sphere with $q = 22.24159$ at 4 loss levels ($l_0 = 35$) in comparison with a normal sphere with a neighbouring size parameter of $q = 22.14159$. Figure 4 (a) and (d) referring to the respective factors of $a_l$ and $c_l$ demonstrate similar curves for the super-enhancement and normal spheres apart from the positions of their amplitude peaks. An additional peak of $b_l$ curve (pink window) is found at the position of order mode of $l = 27$ in Figure 4 (b) i for the super-enhancement sphere in comparison with a flat line shown at the same position for the normal sphere in Figure 4 (b) ii. The dramatic distinction happens in the internal electric field regarding the factor of $d_l$ in Figure 4 (c). A sharp rise reaching the amplitude of 250 is at the order mode of $l = 27$ for the curves of the super-enhancement sphere in Figure 4 (c) i, by contrast, the maximum amplitude of the normal sphere is less than 3.5 in Figure 4 (c) ii. This indicates that the super-enhancement effect in the sphere with $q = 22.24159$ is driven by the factor of $d_l$ for the scattering at the order mode of $l = 27$ in the internal electric field. Also, the scattering amplitudes of $a_l$, $b_l$, $c_l$, and $d_l$ against order mode – $l$ are illustrated in Figure 5 for analysis of the super-enhancement effect in another sphere with $q = 28.64159$ in comparison with its neighbouring normal sphere with $q = 28.74159$ ($l_0 = 43$). Several extra amplitude peaks can be seen on the curves of the super-enhancement sphere in Figure 5, including those at $l = 35$ on $a_l$ curves in Figure 5 (a), $l = 31$ on $b_l$ curves in Figure 5 (b), $l = 31$ on $d_l$ curves in Figure 5 (c), and $l = 35$ on $c_l$ curves in Figure 5 (d). It should be pointed out that the factor of $c_l$ representing the scattering in the internal magnetic field dominates the contributions to the super-enhancement effect in this sphere because of an amplitude peak approximately achieving 1500 at the order mode of $l = 35$, as shown in Figure 5 (d).

Apart from above two examples, the identical analyses of scattering amplitudes were applied to all super-enhancement spheres manifested in Figure 1 and validated that the super-enhancement effect is driven by the scattering of a single order mode, while it can be stimulated in either internal electric field or internal magnetic field. It is consistent with the field intensity distributions of the super-enhancement spheres because all polar hotspots are inside the sphere. Based on the statistical data of these analyses, it is noted that only two extreme enhancements (maximum $|E^2|$ field intensity of 3950 at $q = 28.64159$ and 2952 at $q = 46.54159$ respectively) are stimulated by the internal magnetic fields of the spheres, and the others are all from internal electric fields. This could conclude that the super-enhancement focusing effect induced by internal magnetic field is rare in Teflon spheres, but able to provide much stronger enhancement than that for electric field super-enhancements. Meanwhile, all amplitude peaks at the super-enhancement order modes shown in Figure 4 and 5 are highly sensitive to the loss increase and only exist at the loss levels of $k = 0$ and 1.40e-4. It could be related to the field intensity decline of the hotspots at the top pole of the sphere with an increase of Teflon loss, as shown in Figure 2 and 3, moreover proves that a tiny growth of loss magnitude can adequately offset this super-enhancement effect through the inhibition of scattering from the particular order modes.

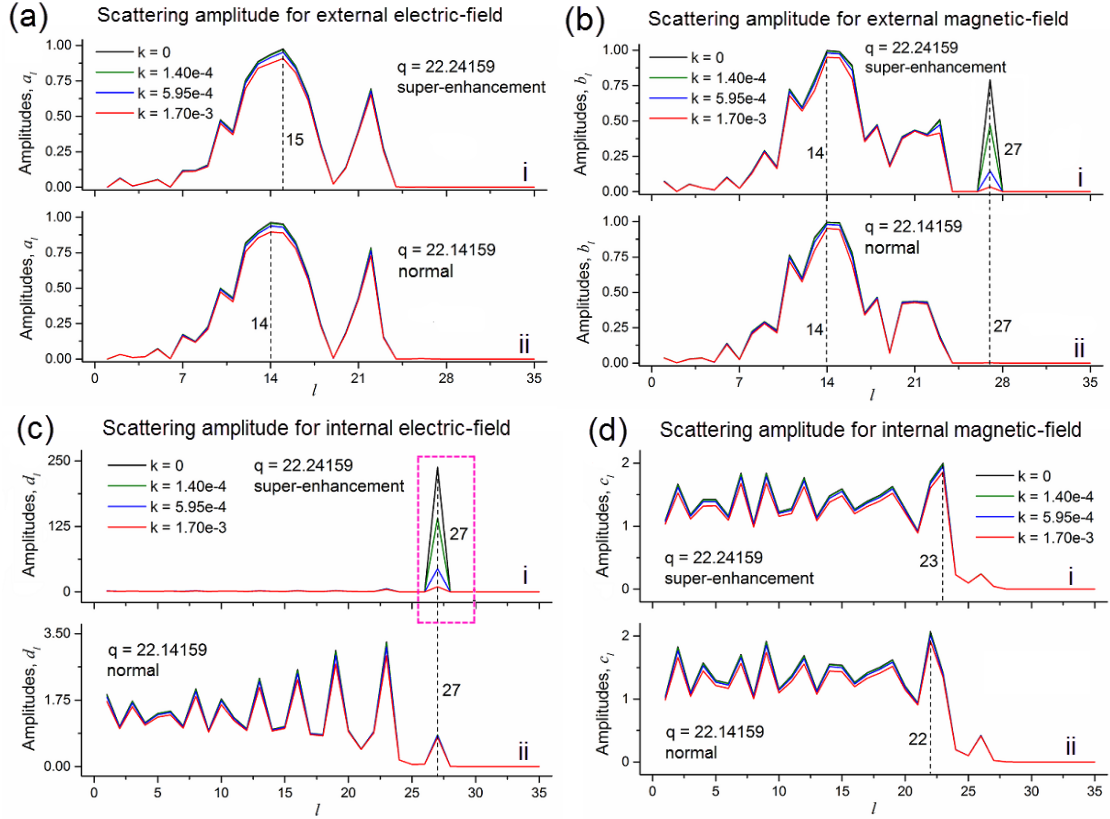

Figure 4. Scattering amplitudes for factors of $a_l$, $b_l$, $c_l$, and $d_l$ against order mode – $l$ for the spheres with $q = 22.24159$ and $q = 22.14159$ on 4 loss levels

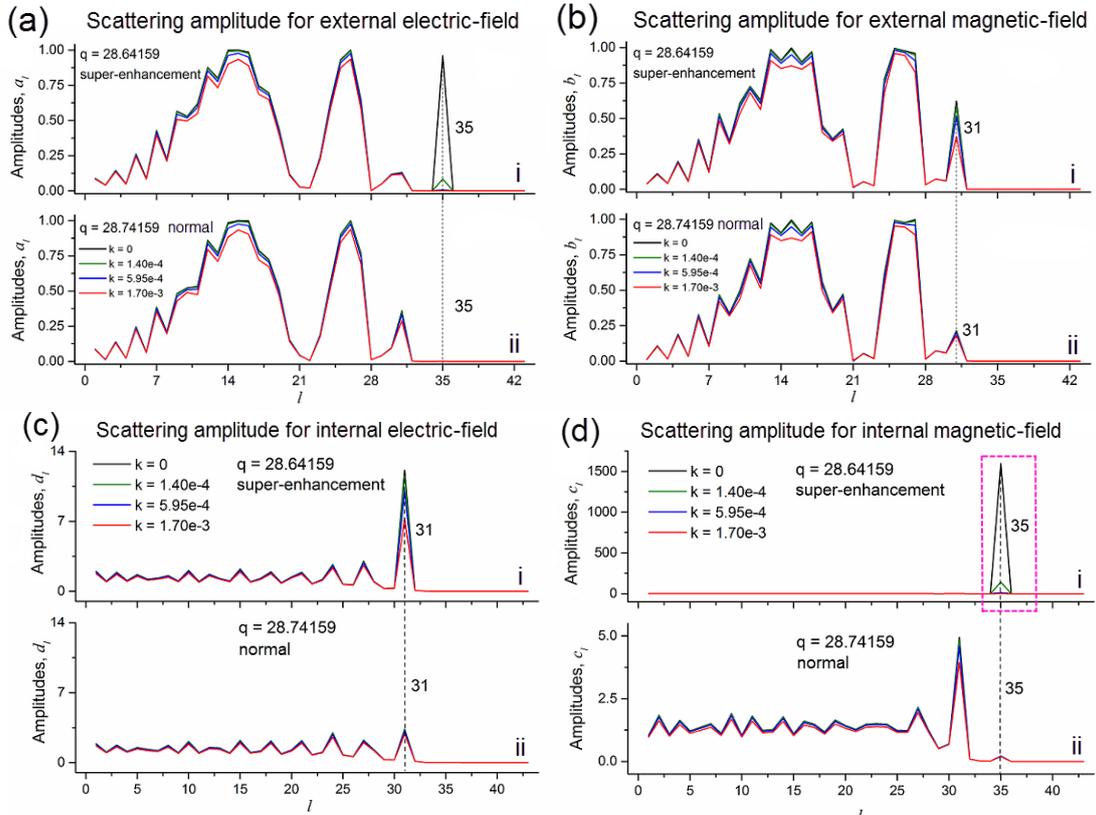

Figure 5. Scattering amplitudes for factors of $a_l$, $b_l$, $c_l$, and $d_l$ against order mode – $l$ for the spheres with $q = 28.64159$ and $q = 28.74159$ on 4 loss levels

## 3.3 Transverse dimension under loss impact

The distributions of $|E^2|$ field intensity in TE mode are inclined to illustrate more features of Whispering-gallery mode in the super-enhancement spheres, which reflects on the multiple elliptical hotspots around a sphere pole at the loss levels of $k = 0$ and 1.40e-4. The elliptical hotspot in TE mode normally has a smaller transverse dimension compared to that in TM mode in a super-enhancement sphere. This characteristic can be found in the comparison between $FWHM_{TE}/a$ and $FWHM_{TM}/a$ in Figure 1 (c) and (d). Also, it is noted that an appropriate increase of Teflon loss is beneficial to reducing the transverse dimension of a focus in the particular super-enhancement sphere, e.g. the sphere with $q = 23.74159$ whose position is marked by a dashed line in Figure 1 (c). Its $FWHM_{TE}/a$ at the loss level of $k = 1.40e-4$ is much smaller than that for lossless Teflon with $k = 0$ and the diffraction limit. These focus profiles of $|E^2|$ field intensities in TE mode are plotted for that sphere at the different loss levels to analyse this phenomenon in Figure 6. Here multiple peaks can be found in both profiles of $k = 0$ and 1.40e-4 at the focal position of $z/a = 0.92$. $FWHM_{TE}/a$ of the curve with $k = 0$ (red line on black curve, length of 0.236) extends over three peaks, as shown in Figure 6. However, increase of loss magnitude of Teflon effectively decrease the intensities of side peaks due to the subsequent stronger attenuation within the same transmission distance in the sphere, which leads that only central peak can be covered by $FWHM_{TE}/a$ of the curve with $k = 1.40e-4$ (red line on green curve, length of 0.058) which is shorter than $DL/a$ of 0.093. This could be considered as the main mechanism to cause the corresponding change of transverse dimension in this sphere under the loss impact.

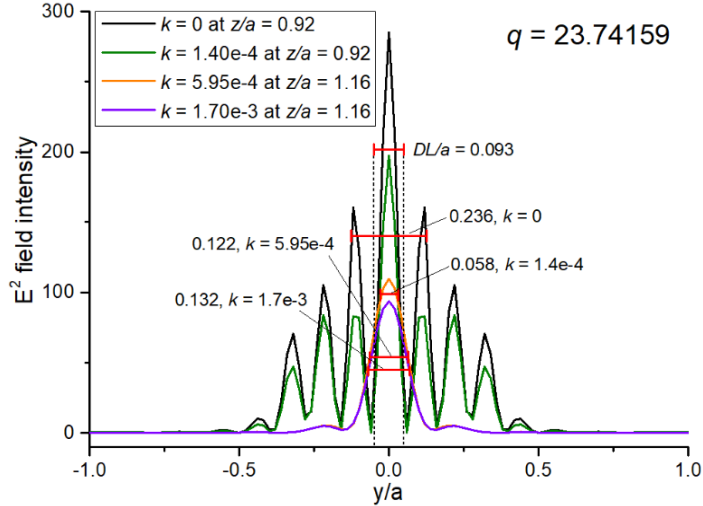

Figure 6. Profile of $|E^2|$ field intensity of the focus of the sphere with the size parameter - $q = 23.74159$ along the *y*-axis

## 4. Conclusions

In this paper, we use an analytical model to systematically investigate the super-enhancement focusing of Teflon spheres with particular size parameters. These super-enhancement Teflon spheres have the unique layout of hotspots at the sphere poles in near-field distributions with a strong field intensity enhancement and great potential of overcoming diffraction limit. An analysis of their scattering amplitudes demonstrates that this super-enhancement effect should be caused by the internal scattering of a single order mode in either electric or magnetic field, but enhancement from internal magnetic field is much stronger. Besides, it was found that this super-enhancement focusing effect can be weakened by the increase of Teflon intrinsic loss but inducing a focus whose size is beyond the diffraction limit and smaller than that for an ideal case of lossless Teflon.

**Acknowledgements**
The authors received the financial supports from the Centre for Photonic Expertise (CPE) with grant No. 81400 by Welsh European Funding Office (WEFO), UK and the European Union H2020 research and innovation programme under grant agreement No. 737164. Part of the work was carried out within the framework of the Tomsk Polytechnic University Competitiveness Enhancement Program, Russia.